\documentstyle[12pt]{article}
\begin{document}
\newcommand{\Dirac}{/\!\!\!\!D}
\newcommand{\beq}{\begin{equation}}
\newcommand{\eeq}[1]{\label{#1}\end{equation}}
\newcommand{\bea}{\begin{eqnarray}}
\newcommand{\eea}[1]{\label{#1}\end{eqnarray}}
\renewcommand{\Re}{\mbox{Re}\,}
\renewcommand{\Im}{\mbox{Im}\,}
\begin{titlepage}
\begin{center}
\hfill hep-th/9508056 CERN-TH/95-217
\vskip .01in \hfill NYU-TH-95/07/02  IFUM/515/FT
\vskip .4in
{\large\bf Heterotic-Type II String Duality and the H-Monopole Problem}
\end{center}
\vskip .4in
\begin{center}
{\large Luciano Girardello}$^a$,
{\large Massimo Porrati}$^b$, and
{\large Alberto Zaffaroni}$^c$
\vskip .1in
$a$ Dipartimento di Fisica, Universit\`a di Milano, via Celoria 16,
20133 Milan, Italy\footnotemark
\footnotetext{and Theory Division, CERN, CH-1211 Geneva 23, Switzerland}
\footnotemark
\footnotetext{e-mail girardello@vaxmi.mi.infn.it}
\vskip .05in
$b$ Department of Physics, NYU, 4 Washington Pl.,
New York, NY 10003, USA\footnotemark
\footnotetext{e-mail porrati@mafalda.physics.nyu.edu}
\vskip .05in
$c$ Centre de Physique Th\'eorique, Ecole Polytechnique,
F-91128 Palaiseau CEDEX,
France\footnotemark
\footnotetext{Laboratoire Propre du CNRS UPR A.0014}
\footnotemark
\footnotetext{e-mail zaffaron@orphee.polytechnique.fr}
\end{center}
\vskip .4in
\begin{center} {\bf ABSTRACT} \end{center}
\begin{quotation}
\noindent
Since T-duality has been proved only perturbatively and most of the heterotic
states map into solitonic, non-perturbative, type II states,
the 6-dimensional string-string duality between the
heterotic string and the type II string is not sufficient
to prove the S-duality of the
former, in terms of the known T-duality of the latter. We nevertheless
show in detail that the perturbative T-duality, together with the
heterotic-type II duality, does imply the existence of heterotic H-monopoles,
with the correct multiplicity and multiplet structure.
This construction is valid at a
generic point in the moduli space of heterotic toroidal compactifications.
\end{quotation}
\vfill
CERN-TH/95-217\\ August 1995
\end{titlepage}
\eject
\noindent
The standard approach to string theory is intrinsically perturbative: one is
given a recipe whereby one computes, say, the $g$-loop contribution to an
S-matrix element in terms of a (super)conformal 2-d field theory on a
genus-$g$ Riemann surface. Naturally, any technique shedding light on the
non-perturbative dynamics of strings is of the utmost importance. One such
technique is based on the
conjecture that some strongly interacting string models can be
rewritten in terms of other, weakly interacting, ``dual,'' string models.

One of the better understood among
string dualities is that between the heterotic string,
compactified to 6 dimensions on a 4-torus $T_4$, and
the type IIA superstring, compactified on $K_3$~\cite{HT,W}. Evidence
supporting this conjecture has been given in refs.~\cite{HT,W,S,HS}.
If this 6-dimensional  heterotic-type II duality holds, it implies various
results.

One of the most important
is that, upon further compactification of both the
heterotic and type II strings to 4 dimensions on a two-torus $T_2$, there
exists a ``duality of dualities''~\cite{D} (see also \cite{SS}) between the
two strings. This property consists in the following: both the heterotic
string, compactified on $T_4\times T_2$, and the type II string,
compactified on $K_3\times T_2$, have $N=4$, $d=4$ supersymmetry.
They are both invariant under a discrete group of
target space dualities (see~\cite{GPR} for a review on this matter).
This group contains the direct product $SL(2,Z)\otimes SL(2,Z)$. The first
$SL(2,Z)$ acts on the complex structure of the torus $T_2$, and it is called
``U-duality.'' The second, called ``T-duality,''
acts by fractional linear transformations on the complex field
$T= B_{56}+i\sqrt{G}$, where $\sqrt{G}$ is the volume of the two-torus and
$B_{56}$ comes from the dimensional reduction on $T_2$ of
the universal antisymmetric tensor of strings. Both theories are also
conjectured to be invariant under a coupling-constant duality, the
``S-duality,'' which also forms an $SL(2,Z)$ group.
As shown in~\cite{D,W}, under heterotic-type II duality, the T- and
S-dualities are interchanged. Thus, one may be tempted to conclude that
S-duality follows automatically from the 6-dimensional heterotic-type II
duality, since T-duality is a well-established, perturbative
symmetry of strings~\footnotemark.
\footnotetext{i.e. a symmetry holding order by order in the string
loop expansion.}

This statement is not correct as it stands: {\em perturbative} T-duality is
not sufficient to prove S-duality in the dual string. One obvious reason is
that, for instance, the type II {\em perturbative} spectrum contains no state
charged under the vectors coming from the Ramond-Ramond sector. These
vectors are mapped by the heterotic-type II duality
into gauge fields in the Cartan subalgebra of the heterotic gauge group
($E_8\otimes E_8$, for instance). Conversely, heterotic states charged under
the gauge group must be mapped by heterotic-type II duality into solitonic
(non-perturbative) states of the type II string. This means that in order to
prove, say, that the heterotic string compactified on $T_4\times T_2$
is S-dual, one
needs to prove that the type II string is T-dual {\em non-perturbatively}.
Thus one has to find the action of T-duality on the non-perturbative,
solitonic spectrum of this string etc. This task is obviously as complicated
as a direct proof of S-duality for the heterotic string.

On the other hand, perturbative T-duality of the type II string can still be
of use in trying to prove S-duality of the heterotic string: one may
discover perturbative states of the type II string, transforming among
themselves under T-duality,
which map under heterotic-type II duality into perturbative, as well as {\em
non-perturbative states} of the heterotic string, transforming among
themselves under S-duality.

The purpose of this paper is to study in detail this scenario, where
the perturbative T-duality of one string gives
non-perturbative information about S-duality on the dual string.
In particular, we will show that the rigorously proved {\em
perturbative} T-duality of the type II string together with the conjectured
heterotic-type II duality in 6 dimensions imply the existence of
H-monopoles~\cite{GH}, with the right multiplicity and super-multiplet
structure.

The paper is organized as follows: first we write the low-energy effective
action of the type II superstring compactified on $K_3\times T_2$, and show
that T-duality is not a manifest symmetry of this action: to prove that two
models related by a T-duality are in fact equivalent, one needs a Poincar\'e
duality involving the field strengths of the vectors coming from the
Ramond-Ramond sector. This fact already shows that perturbative T-duality is
not the whole story for type II strings. Indeed, an equivalence between
theories involving a Poincar\'e duality among two-forms is intrinsically
non-perturbative, when acting on charged fields~\footnotemark.
\footnotetext{The reason is simple: an ``electrically charged'' particle,
coupling to the Ramond-Ramond field strength $F_{\mu\nu}^{R-R}$ with charge
$g_e=\sqrt{4\pi/\Im T}$,
is mapped into a ``magnetically charged'' particle, coupling to
$\tilde{F}_{\mu\nu}^{R-R}
\equiv (1/2)\epsilon^{\mu\nu\rho\sigma}F_{\rho\sigma}^{R-R}$
with charge $g_m=\sqrt{4\pi T\bar{T}/\Im T}$. At $\Re T=0$, this equation
becomes $g_m=4\pi/g_e$, that is, a non-perturbative
equation relating strong coupling to weak coupling.}
Then, we write the effective action of the heterotic string, compactified on
$T_4\times T_2$, and we show in details how the heterotic-type II duality
works. In particular, we show how the S-, T- and U- dualities of
this heterotic compactification relate to the corresponding dualities (S',
T' and U')
of the type II string. After that, we study the perturbative
spectrum of the type II string, and identify the
states related to ``G-poles,'' i.e.
heterotic states charged  with respect to
the vectors $A_\mu^i$, $i=4,5$, which come from the dimensional reduction
of the 6-dimensional metric. The G-poles saturate an appropriate
Bogomol'nyi bound, thus their mass is non-renormalized~\cite{WO}.
These states must be mapped by S-duality into
H-monopoles, i.e. into states magnetically charged under vectors $B_{\mu i}$,
coming from
the dimensional reduction of the 6-dimensional antisymmetric tensor.
We thus come to the last and
main result of the paper: we will show that the {\em
perturbative} spectrum of the type II string contains all the states
corresponding to the H-monopoles of the heterotic string, with the correct
multiplicities. These states are mapped by the perturbative T-duality of the
type II string into the type II partners of the G-poles.
Thus, at least in this case, we can prove that perturbative T-duality and the
6-dimensional string-string duality do provide non-perturbative
information about S-duality.

Let us begin by showing how the T-duality acts on the low-energy effective
action of the type II superstring compactified on $K_3\times T_2$. We start
with the bosonic part of 6-dimensional theory
obtained by compactifying the type II string
on $K_3$,
\bea
& & \int d^6x\big(\sqrt{-G}\big\{e^{-\Phi}\big[R +
G^{\mu\nu}\partial_{\mu}\Phi\partial_{\nu}\Phi - {1\over 12}G^{\mu\mu^\prime}
G^{\nu\nu^\prime}G^{\sigma\sigma^\prime}H_{\mu\nu\sigma}
H_{\mu^\prime\nu^\prime\sigma^\prime}\nonumber\\
&+& {1\over 8}G^{\mu\nu}Tr(\partial_{\mu}\hat M
\hat L\partial_{\nu}\hat M\hat L)\big] -
G^{\mu\mu^\prime}G^{\nu\nu^\prime}F_{\mu\nu}^a
(\hat{L}\hat{M}\hat{L})_{ab}F_{\mu^\prime\nu^\prime}^b\big\}\nonumber\\&-&
{1\over 4}\epsilon^{\mu\nu\sigma\rho\tau\eta}B_{\mu\nu}F_{\sigma\rho}^a
\hat{L}_{ab} F_{\tau\eta}^b\big).
\eea{type}
The 24 Abelian gauge fields, with field strength $F^a_{\mu\nu}$,
come from the reduction of the vector field and the
third-rank antisymmetric field in ten dimensions, both originating in the
Ramond-Ramond sector of the superstring. The
symmetric matrix-valued scalar field $\hat M$
parametrizes the coset $O(4,20)/(O(4)\times 0(20))$ and satisfies
 $\hat M\hat L\hat M = \hat L$,
 where
\beq
\hat L = \left(\begin{array}{ccc}0&I_4&0\\I_4&0&0\\0&0&-I_{16}
\end{array}\right).
\eeq{add1}
Its components come from the reduction of the metric and antisymmetric
 tensor field. It is crucial that
\beq
H_{\mu\nu\sigma} = \partial_{\mu}B_{\nu\sigma} + \mbox{cyclic permutations}.
\eeq{add2}
without Chern-Simons term.\par
In order to perform the dimensional reduction from 6 to 4 dimensions,
it is convenient to introduce tangent
indices, and a parametrization~\cite{schwarz} in which the 6-dimensional
vielbein is
\beq
\hat e_{\hat\mu}^{\hat r} = \left(\begin{array}{cc}e_{\mu}^r&A_{\mu}^iE_i^a\\
0&E^a_j\end{array}\right).
\eeq{add3}
{}From now on, we shall use, whenever necessary, hatted fields and indices
to denote 6-dimensional quantities ($\hat\mu=(\mu,i), i=5,6$ and
$\hat r=(r,a), a=5,6$). Internal indices are raised and lowered by the
metric $h_{ij}=E_i^a\delta_{ab}E_j^b$.\par
We obtain new scalars $h_{ij},B_{ij},A_i^a$ from the internal components of the
6-dimensional metric, antisymmetric tensor and gauge fields. 4 new
vectors, $A_{\mu}^i,B_{\mu i}$, come from the off-diagonal terms of the
metric and antisymmetric tensor.
It is convenient to perform the reduction by starting from the tangent-index
expressions, using the following redefinition
\bea
\hat e_r^{\hat\mu}\hat e^{\hat\nu}_s\hat B_{\hat\mu\hat\nu} &=&
e_r^{\mu}e_s^{\nu}B_{\mu\nu}\nonumber\\
\hat e_r^{\hat\mu}\hat B_{\hat\mu i}&=&e^{\mu}_rB_{\mu i}\nonumber\\
\hat e_r^{\hat\mu}\hat A_{\hat\mu}^a &=& e_r^{\mu}A_{\mu}^a
\eea{add4}
and then convert back to world indices. With the new definitions,
\bea
B_{\mu\nu}&=& \hat B_{\mu\nu} + {1\over 2}(A_{\mu}^iB_{\nu i} -
A_{\nu}^iB_{\mu i}) - A_{\mu}^iB_{ij}A_{\nu}^j\nonumber\\
B_{\mu i} &=& \hat B_{\mu i} + B_{ij}A_{\mu}^j\nonumber\\
A_{\mu}^a &=& \hat A_{\mu}^a -A^a_iA_{\mu}^i.
\eea{add5}
The dimensionally reduced Lagrangian reads:
\bea
&&\int d^4x\big(\sqrt{-g}\big\{e^{-\phi}\big[R +
g^{\mu\nu}\partial_{\mu}\phi\partial_{\nu}\phi - {1\over 12}g^{\mu\mu^\prime}
g^{\nu\nu^\prime}g^{\sigma\sigma^\prime}H_{\mu\nu\sigma}
H_{\mu^\prime\nu^\prime\sigma^\prime}\nonumber\\
&+& {1\over 8}g^{\mu\nu}Tr(\partial_{\mu}{\cal M}J\partial_{\nu}{\cal M}J)
+ {1\over 8}g^{\mu\nu}Tr(\partial_{\mu}\hat M\hat L\partial_{\nu}\hat M\hat
L)\big] + 2\sqrt{h}g^{\mu\nu}h^{ij}\partial_{\mu}A_i^a(\hat L\hat M\hat L)_{ab}
\partial_{\nu}A_j^b\big\}\nonumber\\
&-& \int d^4x\sqrt{-g}\big\{e^{-\phi}\big[{1\over 4}
F_{\mu\nu}^A{\cal M}_{AB}F^{\mu\nu B}\big]
+ \sqrt{h}(F_{\mu\nu}^a+G_{\mu\nu}^iA_i^a)(\hat L\hat M\hat L)_{ab}
(F^{\mu\nu b}+G^{\mu\nu j}A_j^b)\big\}\nonumber\\
&+& \int d^4x\big[ -{1\over 2}B_{56}\epsilon^{\mu\nu\sigma\rho}
 (F_{\mu\nu}^a+G_{\mu\nu}^iA_i^a)\hat L_{ab}
\big(F_{\sigma\rho}^b+G_{\sigma\rho}^jA_j^b\big)
+ \epsilon^{\mu\nu\sigma\rho}\epsilon^{ij}H_{\mu\nu i}A_j^a
\hat L_{ab}(F_{\sigma\rho}^b+{1\over 2}G_{\sigma\rho}^jA_j^b)\nonumber\\
&-&{1\over 3}
\epsilon^{\mu\nu\sigma\rho}H_{\mu\nu\sigma}\epsilon^{ij}A_i^a
\hat L_{ab}\partial_{\rho}A_j^b\big].
\eea{r1}
Here we defined a shifted dilaton field $\phi=\Phi -\log\sqrt{h}$, and
the  curvatures
\beq
F_{\mu\nu}^A = \left(\begin{array}{c}G_{\mu\nu}^i=\partial_{\mu}A^i_{\nu}
-\partial_{\nu}A^i_{\mu}\\H_{\mu\nu i}=\partial_{\mu}B_{i\,\nu}
-\partial_{\nu}B_{i\,\mu}\end{array}\right)\;\;\;A=(i,j).
\eeq{add6}
The kinetic term of vectors is given by the matrices
\beq
{\cal M}=
\left( \begin{array}{cc} h_{ij} - B_{ik}h^{kl}B_{lj} & B_{ik}h^{kj} \\
   -h^{ik}B_{kj} & h^{ij} \end{array} \right),
\;\;\; J = \left(\begin{array}{cc}0&I_2\\I_2&0\end{array}\right).
\eeq{add7}
In the reduction, $H_{\mu\nu\sigma}$ acquired a Chern-Simons term with respect
to the vectors coming from the metric and antisymmetric tensor,
\beq
H_{\mu\nu\sigma} = \partial_{\mu}B_{\nu\sigma}
-{1\over 2}(A_{\mu}^iH_{\nu\sigma i} + B_{\mu i}G^i_{\nu\sigma})
+ \mbox{cyclic permutations}.
\eeq{add8}
{}From string theory we know that the compactification on a two-torus induces
an $O(2,2;Z)$ symmetry of the perturbative spectrum (see~\cite{GPR} and the
discussion of the spectrum below). $O(2,2;Z)$ is split into
two copies of $SL(2;Z)$, called T- and U-duality. Let us show how they act on
 the effective Lagrangian. At first, we must recall that $O(2,2;Z)$ acts on the
 matrix ${\cal M}$~\cite{GPR} in the following way
\beq
O(2,2;Z):\;\;\; {\cal M}\rightarrow \Omega {\cal M}\Omega^T.
\eeq{O2}
The kinetic term for the scalars is obviously invariant,
and so also the kinetic term for the vectors $F^A_{\mu\nu}$,
provided we simultaneously
trasform the field strengths as follows:
\beq
O(2,2;Z):\;\;\; F^A_{\mu\nu}\rightarrow (\Omega^{-1})_{BA}F_{\mu\nu}^B.
\eeq{add9}
The terms containing the gauge fields $F^a_{\mu\nu}$
are more complicated since U and T act in a very different way.
To see this, we must look
more carefully at the action of T and U separately.
A useful parametrization for the two-dimensional matrices $h$ and $B$ is;
\beq
h + iB = \sqrt{h}\left[{1\over U_2}
\left(\begin{array}{cc}U_1^2+U_2^2&U_1\\U_1&1\end{array}\right)\right]
+ iB_{56}\left(\begin{array}{cc}0&1\\-1&0\end{array}\right) = \sqrt{h}\tilde h
 + iB_{56}   .
\epsilon
\eeq{adda}
Introducing the two complex numbers $U=U_1+iU_2$, $T=B_{56}+i\sqrt{h}$,
$O(2,2;Z)$ acts as a copy of the standard $SL(2,Z)$ linear fractional
transformations on both variables,
\bea
&&X\rightarrow {aX+b\over cX+d},\;\;\;
P(X)\rightarrow \omega P(X)\omega^T
\nonumber\\&&X=T,U,\;\;\;
\omega =\left(\begin{array}{cc}a&b\\c&d\end{array}\right),
P(X)={1\over X_1}
\left(\begin{array}{cc}X_1^2+X_2^2&X_1\\X_1&1\end{array}\right) .
\eea{sl}
 Given the following expression
for the matrix ${\cal M}$,
\beq
{\cal M} = \left(\begin{array}{cc}1&0\\0&-\epsilon\end{array}\right)
\left[{1\over T_1}
\left(\begin{array}{cc}T_1^2+T_2^2&T_1\\T_1&1\end{array}\right)\otimes
{1\over U_2}\left(\begin{array}{cc}U_1^2+U_2^2&U_1\\U_1&1\end{array}\right)
\right]\left(\begin{array}{cc}1&0\\0&\epsilon\end{array}\right),
\eeq{add9'}
it is easy to determine the embedding of the T- and U-dualities in the full
group $O(2,2;Z)$. Let us begin with the U-duality. The $O(2,2;Z)$ matrix
corresponding to the $SL(2,Z)$ transformation $\omega$, and the corresponding
transformation of the gauge fields $F^A_{\mu\nu}$, are
\beq
\Omega_U = \left(\begin{array}{cc}\omega &0\\0&(\omega^{-1})^T
\end{array}\right),\;\;\; H_{i\,\mu\nu}\rightarrow
(\omega^{-1})^{ji}H_{j\,\mu\nu},\;
G^i_{\mu\nu}\rightarrow \omega_{ij}G^j_{\mu\nu}.
\eeq{add10}
Imposing the following transformation on the scalars
\beq
A_i^a\rightarrow \omega_{ij}A_j^a
\eeq{addb}
we see that the U-duality is a manifest symmetry of the low-energy effective
Lagrangian.\par
T-duality, on the other hand, is more delicate.
The corresponding $O(2,2;Z)$ transformation is
\beq
\Omega_T =
\left(\begin{array}{cc}a&b\epsilon\\-c\epsilon&d\end{array}\right),
\eeq{add11}
and we see that under the two generators of $SL(2,Z)$ the gauge fields
transform as follows
\beq
T\rightarrow T+1:
\begin{array}{c}G^i_{\mu\nu}\rightarrow G^i_{\mu\nu}\\
H_{i\,\mu\nu}\rightarrow \epsilon_{ij}G^j_{\mu\nu} +
H_{i\, \mu\nu}\end{array}
\;\;\;
T\rightarrow -{1\over T}:
\begin{array}{c}G^i_{\mu\nu}\rightarrow \epsilon^{ij}H_{j\,
\mu\nu}\\H_{i\,\mu\nu}\rightarrow
\epsilon_{ij}G^j_{\mu\nu}\end{array}.
\eeq{slgauge}
Notice that the generator of $T\rightarrow T+1$,
usually realized in a trivial manner, here requires a non-trivial
transformation
of the gauge field $H_{i\,\mu\nu}$,
since it multiplies a term that, in this case, is not
 a topological invariant. It is easy to verify the invariance of the Lagrangian
under the combined transformations~(\ref{add11}) and (\ref{slgauge}),
corresponding to the generator of $T\rightarrow T+1$.\par
For what regards the generator of $T\rightarrow -1/T$,
the kinetic term for scalars
involves only the matrix $h_{ij}$ so it is trivially invariant if the
 scalars do not transform. However, the presence of a factor of $\sqrt{h}$ in
front of the kinetic term for the 24 gauge fields $F^a_{\mu\nu}$,
implies that the T-duality
cannot be realized as a symmetry of the Lagrangian, but it must
involve a Poincar\'e duality on $F^a_{\mu\nu}$.
Equivalently, T-duality is realized only on the equations of motion.
This is most easily seen by adding to the action
the Lagrange multiplier
\beq
\int d^4 x \epsilon^{\mu\nu\sigma\rho}C_{\mu\nu}^aF_{\sigma\rho}^a,
\eeq{add12}
which enforces the Bianchi identities for $F^a_{\mu\nu}$.
Notice that now $F^a_{\mu\nu}$ is an independent variable
and appears only polynomially in the Lagrangian.
If we perform a duality transformation on the modulus $T$ and the gauge fields
$F^A_{\mu\nu}$,
the action is obviously not invariant, but when re-expressed in terms
of the dual gauge field $C^a_{\mu\nu}$, using the $F^a_{\mu\nu}$
equations of motion,
\beq
F^a_{\mu\nu} = \epsilon^{ij}H_{\mu\nu i}A_j^a - T_1(C_{\mu\nu}^a + G_{\mu\nu}^i
A_i^a) + T_2(\hat M\hat L)^{ab}(\tilde C_{\mu\nu}^b + \tilde
G_{\mu\nu}^iA_i^b),
\eeq{poincare}
the action reacquires the
original form.
We have used the following convention for the dual gauge field
\beq
\tilde F^{\mu\nu}={1\over 2\sqrt{-g}}\epsilon^{\mu\nu\sigma\rho}F_{\sigma\rho}.
\eeq{tilde}
We see that an $SL(2,Z)$ transformation on $T$, combined with
the explicit rotation~(\ref{slgauge}) on the gauge fields $F^A_{\mu\nu}$,
and the Poincar\'e duality~(\ref{poincare}) on the gauge fields $F^a_{\mu\nu}$,
is a symmetry of the Lagrangian.
This is how the T-duality is realized on the
low-energy effective action. The equations of motion are of course invariant.
 Let us collect
for further reference the transformations of the gauge
fields
\bea
T\rightarrow T+1 &:&
\begin{array}{c}G^i_{\mu\nu}\rightarrow G^i_{\mu\nu}\\H_{i\,\mu\nu}
\rightarrow \epsilon_{ij}G^j_{\mu\nu} +
H_{i\, \mu\nu}\\F^a_{\mu\nu}\rightarrow F^a_{\mu\nu}\end{array} \nonumber \\
T \rightarrow -{1\over T} &:&
\begin{array}{c}G^i_{\mu\nu}\rightarrow
\epsilon^{ij}H_{j\,\mu\nu}\\H_{i\,\mu\nu}\rightarrow
\epsilon_{ij}G^j_{\mu\nu}\\
F^a_{\mu\nu}\rightarrow \epsilon^{ij}H_{i\,\mu\nu}A_j^a - T_1(F^a_{\mu\nu} +
G^i_{\mu\nu} A_i^a) + T_2(\hat M\hat L)^{ab}(\tilde F^b_{\mu\nu} +
\tilde G^i_{\mu\nu}A_i^b)
\end{array}
\eea{tduality}

Next, we must examine the duality between heterotic and type II strings.
The equivalence we need is between the heterotic string theory compactified
to 6 dimensions on
a 4-torus, and  the type II superstring compactified
on $K_3$~\cite{HT,W,S,HS}. At the level
of low-energy Lagrangians, the string-string duality is
realized by the following redefinitions of the 6-dimensional fields,
\beq
\Phi^\prime = -\Phi,\;\;\; G_{\mu\nu}^\prime = e^{-\Phi}G_{\mu\nu},
\;\;\;
\sqrt{-G^\prime}e^{-\Phi^\prime}H^{\prime\mu\nu\sigma} =
{1\over 6}\epsilon^{\mu\nu\rho\sigma\tau\eta}H_{\sigma\tau\eta}.
\eeq{dual}
This redefinition
maps the equations of motion of the type II Lagrangian~(\ref{type})
into the equations of motion of the heterotic Lagrangian
\bea
&& \int d^6x\sqrt{-G^\prime}e^{-\phi^\prime}\big[R^\prime +
G^{\prime\mu\nu}\partial_{\mu}\Phi^\prime\partial_{\nu}\Phi^\prime - {1\over
12}G^{\prime\mu\mu^\prime}
G^{\prime\nu\nu^\prime}G^{\prime\sigma\sigma^\prime}H^{\prime}_{\mu\nu\sigma}
H^{\prime}_{\mu^\prime\nu^\prime\sigma^\prime}\nonumber\\
&& + {1\over 8}G^{\prime\mu\nu}Tr(\partial_{\mu}\hat M^{\prime}
\hat L\partial_{\nu}\hat M^{\prime}\hat L) -
G^{\prime\mu\mu^\prime}G^{\prime\nu\nu^\prime}F^{\prime a}_{\mu\nu}
(\hat{L}\hat{M}^{\prime}\hat{L})_{ab}F^{\prime b}_{\mu^\prime\nu^\prime}\big],
\eea{heterotic}
where
\beq
H^\prime_{\mu\nu\rho} = \partial_{\mu}B^\prime_{\nu\rho} -
 2A^{\prime a}_{\mu}\hat L_{ab}F^{\prime b}_{\nu\rho}
+ \mbox{cyclic permutations}.
\eeq{add13}
Notice that the crucial ingredient of the redefinition is a Poincar\'e
duality on the third-rank form $H$ in 6 dimensions.
The equivalence of the two
models can be seen also by adding to the type II action~(\ref{type}) a
Lagrange multiplier
\beq
\int d^6 x
\epsilon^{\mu\nu\rho\sigma\tau\eta}
H_{\mu\nu\rho}\tilde H^{\prime}_{\sigma\tau\eta}.
\eeq{add14}
Trading $H$ for $\tilde H^{\prime}$,
and using the equations of motion, we recover
(after a Weyl rescaling, a change of sign of the dilaton, and defining
$H^\prime$ as $\tilde H^\prime$ plus the Chern-Simons contribution) the
heterotic Lagrangian~(\ref{heterotic}).\par
Compactifying further to 4 dimensions on a two-torus, using the same field
redefinitions of the type II case, and paying attention to
the Chern-Simons term, we obtain the reduced Lagrangian (where the primes are
suppressed for simplicity)
\bea
&&\int d^4x\sqrt{-g}e^{-\phi}\big[R +
g^{\mu\nu}\partial_{\mu}\phi\partial_{\nu}\phi - {1\over 12}g^{\mu\mu^\prime}
g^{\nu\nu^\prime}g^{\sigma\sigma^\prime}H_{\mu\nu\sigma}
H_{\mu^\prime\nu^\prime\sigma^\prime}\nonumber\\
&&+ {1\over 8}g^{\mu\nu}Tr(\partial_{\mu}{\cal N}L\partial_{\nu}{\cal N}L)
- {1\over 4}
F_{\mu\nu}^{\alpha}{\cal N}_{\alpha\beta}F^{\mu\nu\beta}\big],
\eea{add14'}
where we have defined
\beq
F^{\alpha}_{\mu\nu} = \left(\begin{array}{c}G^i_{\mu\nu}\\H_{i\,\mu\nu}\\
2F^a_{\mu\nu}\end{array}\right),\;\;\;
L = \left(\begin{array}{ccc}0&I_6&0\\I_6&0&0\\0&0&-I_{16}\end{array}\right).
\eeq{add15}
Here, $H_{\mu\nu\tau}$ has acquired a full Chern-Simons term
\beq
H_{\mu\nu\rho} = \partial_{\mu}B_{\nu\rho} -
{1\over 2}A_{\mu}^{\alpha}\hat L_{\alpha\beta}F^{\beta}_{\nu\rho}
+ \mbox{cyclic permutations}.
\eeq{add16}
The scalar matrix ${\cal N}$ is expressed in terms of the original fields of
the theory as follows:
\beq
\left(\begin{array}{ccc}h+(C-B)h^{-1}(C+B)+A(\hat L\hat M\hat L)A&
-(C-B)h^{-1}&(C-B)h^{-1}(A\hat L)+A(\hat L\hat M\hat L)\\
-h^{-1}(C+B)&h^{-1}&-h^{-1}(A\hat L)\\
(\hat LA)h^{-1}(C+B)+(\hat L\hat M\hat L)A
&-(\hat LA)h^{-1}&(\hat LA)h^{-1}(A\hat L)+(\hat L\hat
M\hat L)\end{array}\right),
\eeq{add17}
where
\beq
C_{ij}={1\over 2}A_i^a\hat L_{ab}A_j^b.
\eeq{add18}
The heterotic 4-dimensional string also
possesses an S-duality~\cite{sen}, which
acts on the complex coupling constant,
\beq
\lambda = \psi + ie^{-\phi},\;\;\; H^{\mu\nu\rho}=-(\sqrt{-g})^{-1}e^{\phi}
\epsilon^{\mu\nu\rho\sigma}\partial_{\sigma}\psi,
\eeq{add19}
as a linear fractional transformation, and on the gauge fields as the
Poincar\'e duality
\beq
\lambda\rightarrow {a\lambda+b\over c\lambda+d}\;\;\;
F_{\mu\nu}^a\rightarrow (c\lambda_1+d)F_{\mu\nu}^a-c\lambda_2(L{\cal N})_{ab}
\tilde F_{\mu\nu}^a .
\eeq{sduality}

The  type II and heterotic 4-dimensional Lagrangians are obviously, but
not manifestly, equivalent. When reduced to 4 dimensions,
the simple 6-dimensional redefinition~(\ref{dual}) becomes
a less obvious redefinition (Poincar\'e duality)
of $H_{\mu\nu\tau},G_{\mu\nu},G_{ij},B_{ij}$.
In particular, under string-string duality,
the T-duality of the type II superstring is
mapped to the S-duality of the heterotic string. In fact, by reducing
eq.~(\ref{dual}) to 4 dimensions (taking into account the
Chern-Simons term in the heterotic side), we learn that
\beq
B_{56}\rightarrow \lambda_1^\prime\;\;\; {1\over 6}\epsilon_{\mu\nu\tau\sigma}
H_{\nu\tau\sigma}\rightarrow \sqrt{-g^\prime}e^{-\phi^\prime}
\partial^{\mu}B^{\prime}_{56}.
\eeq{add20}
Thus,
if we exchange the role of $\sqrt{h}$ and $e^{-\phi}$ after the
redefinition~(\ref{add20}),
the $T$ modulus is mapped in the complex coupling constant $\lambda^\prime$
and vice-versa. The redefinition of the gauge field $H_i$ reads
\beq
H_{\mu\nu i}^\prime - G_{\mu\nu}^j(C_{ij}+B_{ij}^\prime ) - A_i^a\hat
L_{ab}F^b_{\mu\nu} = {e^{-\phi}\over (\sqrt{h})^2}h_{ij}\epsilon^{jk}(\tilde
H_{\mu\nu k}
-\tilde G_{\mu\nu}^tB_{kt}).
\eeq{red}
It is now a simple, though tedious, exercise to check that the T-duality
transformations~(\ref{tduality}) for the type II gauge fields
$(G^i_{\mu\nu},H_{i\,\mu\nu},F^a_{\mu\nu})$,
when re-expressed in terms of the primed variables, reproduce exactly the
S-duality transformations~(\ref{sduality}) of the heterotic gauge fields
$(G^{\prime i}_{\mu\nu},H_{i\,\mu\nu}^\prime ,F^{\prime a}_{\mu\nu})$.

Let us now turn to the study of the type II superstring (perturbative)
spectrum. We want to
find states that may correspond to massive, short multiplets of the $N=4$
supersymmetry algebra~\cite{FS}. These multiplets have the same number of
components as a ``long'' $N=2$ multiplet, and their common supersymmetric mass
saturates a Bogomol'nyi bound. This means that their squared mass is
proportional to the sum of squares of some Abelian (central) charges
$Q_e^I$, $I=1,..,6$, and a constant matrix $M_{IJ}$:
\beq
m^2_{\mbox{short multiplet}}=Q_e^IM_{IJ}Q_e^J.
\eeq{m1}
The charges $Q_e^I$ are ``electric,'' because no perturbative state can have
a non-zero ``magnetic'' charge. To identify these charges, and find which
type II massless vector they correspond to, we must recall some elementary
facts about superstring theory~\cite{Sch}.

The spectrum of the type II superstring compactified on $K_3\times T_2$ can
be easily written in the light-cone gauge, and in the Ramond-Neveu-Schwarz
formalism.
We will work at a generical point of the separate moduli space of $K_3$ and
$T_2$, but with a compactification for which $A_i^a=0$, since otherwise the
corresponding conformal field theory is no longer constructed with free fields
and involves non-trivial R-R deformations.
 We need only consider the space-time bosons, since fermions can
be obtained from them by space-time supersymmetry transformations. Since one
can choose independently the boundary conditions of the left- and
right-moving world-sheet fermions, as either antiperiodic (Neveu-Schwarz
b.c. or NS) or periodic (Ramond b.c. or R), one obtains 4 sectors in the
Hilbert space of string states, denoted as usual by NS-NS, NS-R, R-NS or R-R.
The space-time bosons of the type II string arise from both the NS-NS and
the R-R sectors. Their mass is given by the standard light-cone formula
($\alpha'=1/2$):
\beq
{1\over 2} m^2 = N_R^{ST} + L_0^{T_2} + L_0^{K_3} -{1\over 2}=
N_L^{ST} + \bar{L}_0^{T_2} + \bar{L}_0^{K_3}-{1\over 2}.
\eeq{m2}
Here  $N_R^{ST}$ ($N_L^{ST}$) is the usual right (left) transverse
space-time oscillator number. In detail, we have two free bosons
(the transverse space-time coordinates $X^1(\sigma,\tau)$, $X^2(\sigma,\tau)$)
together with their world-sheet supersymmetric partners, the Neveu-Schwarz
fermions $\psi_R^\mu(\sigma-\tau)$, $\psi_L^\mu(\sigma+\tau)$, $\mu=1,2$.
Their mode expansion is
\bea
X^\mu(\sigma,\tau)&=&  x^\mu + k^\mu\tau +
{i\over \sqrt{2}} \sum_{n\neq 0} {1\over n} [\alpha_{R\, n}^\mu
e^{-in(\tau-\sigma)} + \alpha_{L\, n}^\mu e^{-in(\tau +\sigma)}], \nonumber \\
\psi_R^\mu(\sigma-\tau) &=& \sum_{r\in Z+a/2} \psi_{R\,
r}^\mu e^{-ir(\tau-\sigma)}, \;\;\;
\psi_L^\mu(\sigma+\tau) =\sum_{r\in Z+a/2} \psi_{L\,
r}^\mu e^{-ir(\tau+\sigma)}, \nonumber \\
{[}\alpha_{R\,n}^\mu, \alpha_{R\,m}^\nu ] &=& [ \alpha_{R\,n}^\mu,
\alpha_{R\,m}^\nu ]
= m\delta^{\mu \nu}\delta_{n+m,0} , \;\;\;
[ \alpha_{R\,n}^\mu, \alpha_{L\,m}^\nu ]=0, \nonumber \\
\{ \psi_{R,\, r}^\mu, \psi_{R\, p}^\nu \}_+ &=&
\{ \psi_{L,\, r}^\mu, \psi_{L\, p}^\nu \}_+ = \delta^{\mu \nu}
\delta_{q+p,0}, \;\;\;
\{ \psi_{R,\, r}^\mu, \psi_{L\, p}^\nu \}_+ =0.
\eea{m2a}
Here $a$ is a constant equal to 1 in the NS sector and to 0 in the R
sector, while $k^\mu$ is the transverse space-time momentum. The oscillator
number $N_R$ reads
\beq
N_R^{ST} = \sum_{n>0} \alpha_{R\,-n}^\mu \alpha_{R\,n}^\mu + \sum_{r\in N+a/2}
r\psi_{R\,-r}^\mu \psi_{R\, r}^\mu + {1-a \over 8}.
\eeq{m2b}
The formula for $N_L$ is obtained by replacing $R$ with $L$ throughout this
equation.

$L_0^{T_2}$ ($\bar{L}_0^{T_2}$) is the right (left) Virasoro operator of two
free bosons $X^i(\sigma,\tau)$, $i=4,5$,
compactified on a two-dimensional torus ($X^i \approx X^i + 2\pi n$, $n\in
Z$), together with their fermionic right- and left-moving superpartners
$\psi_R^i(\sigma-\tau)$, $\psi_L^i(\sigma + \tau)$.
Their mode expansion is~\cite{NSW,GPR}
\bea
X^i(\sigma,\tau)&=&  x^i + m^i\sigma +h^{ij}(n_j -B_{jk}m^k)\tau +
{i\over \sqrt{2}} \sum_{n\neq 0} {1\over n} [\alpha_{R\, n}^i
e^{-in(\tau-\sigma)} + \alpha_{L\, n}^i e^{-in(\tau +\sigma)}], \nonumber \\
\psi_R^i(\sigma-\tau) &=& \sum_{r\in Z+a/2} \psi_{R\,
r}^ie^{-ir(\tau-\sigma)}, \;\;\;
\psi_L^i(\sigma+\tau) = \sum_{r\in Z+a/2} \psi_{L\,
r}^ie^{-ir(\tau+\sigma)}, \nonumber \\
{[}\alpha_{R\,n}^i, \alpha_{R\,m}^j ] &=& [ \alpha_{R\,n}^i, \alpha_{R\,m}^j ]
= mh^{ij}\delta_{n+m,0} , \;\;\;
[ \alpha_{R\,n}^i, \alpha_{L\,m}^j ]=0, \nonumber \\
\{ \psi_{R,\, r}^i, \psi_{R\, p}^j \}_+ &=&
\{ \psi_{L,\, r}^i, \psi_{L\, p}^j \}_+ = h^{ij} \delta_{q+p,0}, \;\;\;
\{ \psi_{R,\, r}^i, \psi_{L\, p}^j \}_+ =0.
\eea{m3}
$h_{ij}$ and $B_{ij}$ are, respectively, the metric and
antisymmetric tensor of the two-torus.
The integers $m^i\in Z$ are the winding numbers, while the $n_i \in Z$ are
the momenta. Thanks to eq.~(\ref{m3}), the Virasoro operators on the
two-torus read
\bea
L_0^{T_2} + \bar{L}_0^{T_2} &=& {1\over 2} Z^t {\cal M} Z +
N_R^B + N_R^F + N_L^B
 + N_L^F + {1-a\over 4}, \nonumber \\
L_0^{T_2} -\bar{L}_0^{T_2} &=& m^i n_i + N_R^B + N_R^F - N_L^B - N_L^F,
\nonumber \\
N_R^B &=& \sum_{n>0} h_{ij}\alpha_{R\, -n}^i \alpha_{R\, n}^j
\;\;\; N_R^F = \sum_{r\in N + a/2} r \psi_{R\, -r}\psi_{R\, r}, \nonumber \\
N_L^B &=& \sum_{n>0} h_{ij}\alpha_{L\, -n}^i \alpha_{L\, n}^j
\;\;\; N_L^F = \sum_{r\in N + a/2} r \psi_{L\, -r}\psi_{L\, r}.
\eea{m4}
The matrix ${\cal M}$ is the same as the one we
encountered in the dimensional reduction of
the low-energy effective action, and $Z$ is a column vector:
\beq
Z= \left( \begin{array}{c} m^i \\ n_i \end{array} \right).
\eeq{m5}
{}From eq.~(\ref{m3}) we learn that the form of the
internal, two-dimensional momenta is
\bea
p_R^i &=& {1\over\sqrt{2}}(m^i - h^{ij}(n_j -B_{jk}m^k))\nonumber\\
p_L^i &=& {1\over\sqrt{2}}(m^i + h^{ij}(n_j -B_{jk}m^k)) .
\eea{momenti}
 The Hamiltonian restricted to zero modes reads:
\beq
H={1\over 2}(p_R^2+p_L^2)\equiv {1\over 2} (p^i_Rp^i_R + p^i_Lp^j_L)h_{ij}.
\eeq{addc}
The compactifications~\cite{GPR} are in one-to-one correspondence
with the even self-dual Lorentzian lattice $\Gamma^{(1,1)}$ spanned by
the vectors $(p_R,p_L)$, which have indeed even-integer Lorentzian norm:
\beq
p_L^2 - p_R^2 = 2n_im^i \in 2Z .
\eeq{addd}
The moduli space of toroidal compactifications is therefore isomorphic to
$O(2,2;R)/(O(2)\times O(2))$. The spectrum is known to be invariant under
the T-duality transformation $O(2,2;Z)$, which acts as a linear transformation
on $Z$: $Z\rightarrow \omega Z$, $\omega\in SL(2,Z)$.

The superstring coordinates, compactified on $K_3$, describe an $N=4$
two-dimensional superconformal field theory. For our purposes, we do not
need the complete spectrum of the theory, but only the lowest
conformal-weight states in the NS and R sectors.
All unitary representations of the
$N=4$ superconformal algebra have been classified in~\cite{EOT}. They are
labelled by the conformal weight, $h$, and by an internal $SU(2)$ spin $l$.
For central charge $c=6$ (recall that
$K_3$ has real dimension 4), and since we have two
superconformal $N=4$ algebras, with Virasoro operators $L_0$ and
$\bar{L}_0$, respectively, the generic $K_3$ state belongs to an  irreducible
representation of $SU(2)\otimes SU(2)$, and reads:
\beq
|h,l,\bar{h},\bar{l}\rangle;\;\;\; L_0|h,l,\bar{h},\bar{l}\rangle=
h|h,l,\bar{h},\bar{l}\rangle, \;\;\; \bar{L}_0 |h,l,\bar{h},\bar{l}\rangle
= \bar{h} |h,l,\bar{h},\bar{l}\rangle .
\eeq{m6}
As shown in ref.~\cite{Sei}, the lowest-weight states of the NS-NS sector
are: a)
the (unique) $SL(2,R)$-invariant vacuum $|0,0,0,0\rangle$. b) 20
4-vectors (representations $(1/2,1/2)$ of $SU(2)\otimes SU(2)$) with
conformal weights $h=\bar{h}=1/2$: $|1/2,1/2,1/2,1/2\rangle$. The vacuum
$|0,0,0,0\rangle$ is even under the standard GSO projection~\cite{GSO} of
the type IIA superstring, while
the (20)
states $|1/2,1/2,1/2,1/2\rangle$ are odd. In the R-R sector, the lowest
conformal-weight states are obtained by the spectral flow from the above NS-NS
states~\cite{Sei}. In particular, the spectral flow of the vacuum gives rise
to a
R-R state $|1/4,1/2,1/4,1/2\rangle$ (a 4-vector of $SU(2)\otimes SU(2)$),
while the spectral flow of each 4-vector $|1/2,1/2,1/2,1/2\rangle$ gives
rise to a singlet $|1/4,0,1/4,0\rangle$. Thus, in total, in the R-R sector,
we find 24 states with conformal weights $h=1/4$, $\bar{h}=1/4$.

The Hilbert space of the
conformal field theory corresponding to the compactification on
$T_2\times K_3$ is just the direct product of the Hilbert spaces of the
theories on the transverse space-time coordinates $T_2$, and $K_3$. The
physical states are odd under the GSO projection. As we are interested only
in states saturating a Bogomol'nyi bound, we must
find the charges $Q_e^I$ of our perturbative type II string states.
This is easily done by noticing that the 4-dimensional Abelian vectors,
which come from the dimensional reduction of the
6-dimensional theory on $T_2$, are,
using our previous notations, $\hat{B}_{\mu i}=B_{\mu i}+A_\mu^jB_{ij}$ and
$\hat{A}_{\mu i}=A_\mu^jh_{ij}$.
The light-cone vertices corresponding to the fields $A_\mu^i$, $B_{\mu i}$
read, at zero 4-dimensional momentum,
\bea
V(B_{\mu i}) &=& \int d\tau d\sigma
(\partial_\sigma X^\mu \partial_\tau X^i
-\partial_\tau X^\mu \partial_\sigma X^i), \nonumber \\
V(A_\mu^i) &=& \int d\tau d\sigma[\partial_\sigma X^\mu (h_{ij}
\partial_\sigma X^j + B_{ij}\partial_\tau X^j)
-\partial_\tau X^\mu (h_{ij}\partial_\tau X^j + B_{ij}\partial_\sigma X^j)] .
\eea{m7}
{}From these equations, one can extract the charges $Q_i$, associated with
$A_\mu^i $, and $\tilde{Q}^i$, associated with $B_{\mu i}$:
\beq
Q_i=\int d\sigma (h_{ij}\partial_\tau X^j  + B_{ij}\partial_\sigma
X^j)=n_i,\;\;\;
\tilde{Q}^i=\int d\sigma \partial_\sigma X^i=m^i.
\eeq{m8}
Let us note that these charges are not those that appear in the asymptotic
expression of the gauge fields, because of the non-canonical kinetic terms
of the vectors $A^A_{\mu}$. If we add a source $\int d^4x
\sqrt{-g}A^A_{\mu}J_A^{\mu}$ to the Lagrangian~(\ref{r1}), we learn
from the equations of motion that, for large $r$ (and $A_i^a=0$),
\beq
F^A_{0i}\rightarrow {q^A\over r^2},\;\;\; q^A={1\over e^{-\phi}}
({\cal M}^{-1})^{AB}
Q_B,\;\;\; Q_A = \left(\begin{array}{c}Q_i \\ \tilde Q^i\end{array}\right).
\eeq{add21}

By comparing eq.~(\ref{m1}) with the mass formula eq.~(\ref{m2}), and using
the explicit formulae just derived above, we find that the Bogomol'nyi
bound is saturated only when either
$N_R^{ST} + N_R^F + N_R^B + h = (1+a)/4$ (no constraint on the left-moving
oscillator numbers and the conformal weight on $K_3$), or
$N_L^{ST} + N_L^F + N_L^B + \bar{h} = (1+a)/4$ (no constraint on the
right-movers) and explicitely reads:
\beq
m^2 = {1\over 2}Z^t({\cal M}+J)Z
= p_R^2 \;\;\mbox{or }p_L^2.
\eeq{add22}

Let us now turn to the heterotic string compactified on $T^6$.
The supersymmetric right-movers are two transverse space-time and
6 internal free bosons, together with their fermionic superpartners. The
non-supersymmetric left-movers are two space-time and 6 internal free
bosons coming from the compactification on $T^4\times T^2$, and
16 free bosons in the Cartan subalgebra of the gauge group $E_8\times E_8$.
The conventions and the normalizations for oscillators are the same as for
the type II string, so let us exhibit only the difference in the zero modes
of the bosons
\bea
X^i_R &=& x^i_R + \sqrt{2}p_R^i(\sigma -\tau) + \mbox{oscillators},
\;\;\;i=5,...,10
\nonumber\\
X^{i,I}_L &=& x^{i,I}_L + \sqrt{2}p^{i,I}_L(\sigma -\tau)+ \mbox{oscillators},
\;\;\; i=5,...,10,
I=1,...,16.
\eea{add23}
Here the internal momenta depend on the compactification data so
 that the vector $(p_R,p_L)$ belongs to the even self-dual Lorentzian lattice
$\Gamma^{(6,22)}$.
We refer to~\cite{GPR} for the explicit expression of the momenta. Here we
limit ourselves to  two particular cases.
When the background vectors of $E_8\times E_8$ are set to
zero, formula~(\ref{momenti}) is reproduced, with the only difference that the
indices now run from 5 to 10.
When, instead, the winding number is set to zero, we obtain~\cite{NSW}:
\bea
p_L^I&=&e^I\nonumber\\
p_L^i&=&{1\over\sqrt{2}}h^{ij}(n_j+A^I_ie^I)\nonumber\\
p_R^i&=&p_L^i
\eea{mo}
where $e^I$ is an element in the root lattice of $E_8\times E_8$.
 Formulae for the Virasoro operators analogous to those given for the
type II superstring are valid for the heterotic string; here we recall only
the main differences.
Since we are interested in the bosonic part of the
spectrum (the fermionic part follows for space-time supersymmetry), we will
consider only the NS sector. The mass-shell condition now reads
\beq
{1\over 2} m^2 = N_R^{ST} + L_0^{T_6} - {1\over 2}=
N_L^{ST} + \bar{L}_0^{T_6} -1.
\eeq{add24}
The Bogomol'nyi condition is now realized when $N_R^{TOT}=1/2$, with the
left-moving oscillators constrained by
\beq
N_L^{TOT} - 1 = {1\over 2}(p_R^2-p_L^2).
\eeq{add25}
and reads
\beq
m^2 = p_R^2 = {1\over 2}Q^{\alpha}(N+L)_{\alpha\beta}Q^{\beta} ,
\eeq{add26}
where the vector $Q^\alpha$ contains the momentum,
the winding number, and the charges
under $E_8\times E_8$, $Q^\alpha =(n_i,m^i,e^I)$.
The asymptotic expression of the gauge fields are given by
\beq
F^\alpha_{0i}\rightarrow {q^\alpha\over r^2},\;\;\; q^\alpha={1\over e^{-\phi}}
(L{\cal N}L)_{\alpha\beta}
Q^\beta                              .
\eeq{add27}

The point of view of searching evidence for S-duality in the heterotic string
strongly reduces the
number of perturbative string states that we can consider: this is because
our knowledge of the non-perturbative solitonic spectrum of the string is
limited to the solutions of the low-energy action for the massless states.
Thus, we are led
to consider only perturbative string states which can become
light in some compactification limit (as, for instance, the large-radius
limit). Only these states are mapped by S-duality into light solitonic
states, which may be found as solutions of the equations of motion of the
low-energy effective action of the string~\cite{sen}. Therefore, we shall
consider only states with zero winding number, for which a field theory limit
exists. To satisfy the condition $N_R^{TOT}=1/2$, we apply to the vacuum
$|0\rangle$
the 8 left-moving fermionic oscillators $\psi_{R-1/2}^{\mu ,i}$, which give
rise by themselves to
the bosonic part of a 16-dimensional representation of $N=4$
supersymmetry (1 vector and 5 scalars). The simplest state satisfying the
Bogomol'nyi bound has $N_L=0$ and $p_L^2=p_R^2+2$.
This condition can be
realized only with states that are charged under $E_8\times E_8$~(\ref{mo});
however, they are
uncharged with respect to the vectors coming from the metric and
antisymmetric tensor, since they have neither internal
 momentum nor winding number on
the compactification torus ($n_i=m^i=0$).
Under an S-duality transformation, the
$p_L^2=p_R^2+2$ states are mapped into
the well-known BPS monopoles. Since we want to compare the heterotic
with the type II
string in which these states are solitonic, non-perturbative excitations (as
any state charged with respect to $E_8\times E_8$), we will not consider them.
More interesting, from our point of view, are the states with $N_L=1$ and
$p_L^2=p_R^2$. They are charged only with respect to the vectors coming
from the metric; indeed, formula~(\ref{mo}) implies $e^I=0$ and
the winding number is zero by assumption. These states are obtained by
applying
the 24 left-moving bosonic oscillators $X^A_{-1}$  to the vacuum
$|0\rangle$.
By tensoring the Lorentz indices, we get 21 vector representations
and 1 spin-2 representation of $N=4$ supersymmetry.
These states are mapped by S-duality
into solitonic solutions, magnetically charged under the vectors coming from
the antisymmetric tensor; such solutions are known as H-monopoles, and they
are therefore predicted to have multiplicity 21+1~\cite{sen}.
These are the states we want to
find in the type II string. States with $N_L> 1$ are easily seen to
need non-zero winding number and are not expected to
have a field theory limit.

The direct study of the multiplicity of the H-monopoles has not given a
complete answer because of the difficulty in analysing the moduli
space of the solution~\cite{GH}. We want to bypass this problem by
looking for these states in the dual type II string, where the counterparts of
heterotic H-monopoles are perturbative.
Indeed, by using eq.~(\ref{add7}) and (\ref{add21}), and setting $A^a_i=0$,
one can check that the fields
$H_{\mu\nu i}^\prime - G_{\mu\nu}^j(C_{ij}+B_{ij}^\prime ) - A_i^a\hat
L_{ab}F^b_{\mu\nu}$ and
 $h_{ij}\epsilon^{jk}(\tilde H_{\mu\nu k}
-\tilde G_{\mu\nu}^tB_{kt})$, which
appear in eq.~(\ref{red}) have an asymptotical charge that is simply
the winding number of the string state in, respectively, the heterotic and the
type II string.
 With a closer look at the same equation, we learn that the type II string
state corresponding to the heterotic state with $p_L^2=p_R^2$ and $N_L=1$,
is both electrically charged and with zero winding number. We can
therefore search for
it in the perturbative spectrum of the type II string.

In the NS-NS sector the mass formula reads:
\beq
{1\over 2}m^2 = {1\over 2}p_R^2 + h + N_R^{TOT} - {1\over 2} =
{1\over 2}p_L^2 + \bar h + N_L^{TOT} - {1\over 2}
\eeq{wind}
and the
Bogomol'nyi bound can be realized by imposing $N_R=1/2$ or $h=1/2$ (or the
same condition in the left sector). The constraint always reads $p_R^2=
p_L^2$, and implies, using formula~(\ref{wind}), that we are indeed
considering
states with zero winding number and a field theoretical interpretation.
 We find 16 states of the form $\psi_{L-1/2}^{A}
\psi_{R-1/2}^{B}|0\rangle$, $A,B=(\mu,i)$,
and 80 states $|1/2,1/2,1/2,1/2\rangle$.
On the other hand in the R-R sector
the mass formula
\beq
{1\over 2}m^2 = {1\over 2}p_R^2 + h + N_R^{TOT} - {1\over 4} =
{1\over 2}p_L^2 + \bar h + N_L^{TOT} - {1\over 4}
\eeq{add28}
can be realized only by $h=\bar h=1/4$, with the same constraint $p_R^2=
p_L^2$.
The Ramond vacuum of the space-time and $T^2$ fermions (4 left-movers and
4 right-movers) $|s,\alpha\rangle$
is labelled by the space-time helicity ($s=1,0,0,-1$) and
by the $T^2$ ``helicity'' ($\alpha=1,0,0,-1$). The physical
states are obtained by multiplying this vacuum by the 4 states obtained
from the spectral flow of the identity, $|1/4,1/2,1/4,1/2\rangle$, and the 20
states $|1/4,0,1/4,0\rangle$, and projecting over GSO odd states. The GSO
projection leaves a total of $384/4=96$ physical states.
This is exactly the multiplicity of 192 bosonic states found
in the perturbative spectrum of the heterotic string.
The helicities of all these 192 states, given in the table below, arrange
exactly into the bosonic
parts of 21 $N=4$ short vector multiplets and 1 spin-2 short multiplet, as
illustrated in table 1.\\ \begin{center}
\begin{tabular}{|| l | l | l ||} \hline\hline
State  & Helicity & Multiplicity \\ \hline
$\psi_{L-1/2}^\mu \psi_{R-1/2}^\nu |0\rangle$ & $\pm 2$ & 1 \\
                                              &  0    & 2 \\ \hline
$\psi_{L-1/2}^\mu \psi_{R-1/2}^i |0\rangle$   & $\pm1$  & 4 \\
$\psi_{L-1/2}^i \psi_{R-1/2}^\mu |0\rangle$   &     &  \\ \hline
$\psi_{L-1/2}^i \psi_{R-1/2}^j |0\rangle$     & 0     & 4 \\ \hline
$|1/2,1/2,1/2,1/2\rangle$ & 0 & 80 \\ \hline
$|\pm 1, \alpha\rangle \otimes [ |1/4,1/2,1/4,1/2\rangle \oplus
|1/4,0,1/4,0\rangle ]$ & $\pm 1$ & 24 (after GSO) \\ \hline
$|0 ,\alpha\rangle \otimes [ |1/4,1/2,1/4,1/2\rangle \oplus
|1/4,0,1/4,0\rangle ]$ & 0 & 48 (after GSO) \\ \hline
\hline
\end{tabular}\vskip .1in Table 1 \end{center} \vskip .01in
If we now perform a T-duality on such states, we obtain, obviously,
a state with non-zero winding number only.
Note that in  the heterotic string, the H-monopoles exist for
$\lambda_1^\prime=0$~\cite{GH};
translated in the type II vocabulary, this means
$B_{56}=0$. Using once more formula~(\ref{red}) with $A_i^a=0$,
 we get a state with the
correct quantum numbers to be identified with the H-monopole. Its existence
and its correct multiplicity are now guaranteed by the perturbatively proved
T-duality of the type II string.
Clearly, we have found only states charged with respect to the 5 or 6
components of the gauge fields G and H; the other H-monopoles can be obtained
by a T-duality in the heterotic string.

In conclusion, the string-string duality turns out to be useful in
studying the non-perturbative dynamics of the heterotic string.
Obviously, there is a price to be paid: in the type II string, many states,
which are present in the perturbative spectrum of the heterotic string,
appear as solitons. The study of these solitons, which lies beyond the scope
of this paper, would be a most powerful test of the string-string
duality itself.
\vskip .1in
\noindent
{\bf Acknowledgements}
\vskip .1in
\noindent
L.G. and A.Z. would like to thank the Department of Physics of NYU for
its kind hospitality. M.P would like to thank the Aspen Center for Physics,
where part of this work was completed.
L.G. is supported in part by the
Ministero dell' Universit\`a e della Ricerca Scientifica e Tecnologica,
by INFN and by ECC, contracts SCI-CT92-0789 and CHRX-CT92-0035.
M.P. is supported in part by NSF under
grant no. PHY-9318171. A.Z. is supported in part by ECC, Projects
ERBCHGCTT93073, SCI-CT93-0340, CHRX-CT93-0340.

\end{document}